# Soliton-effect self-compressed single-cycle 9.6-W mid-IR pulses from a 21-W OPCPA at 3.25 µm and 160 kHz

Ugaitz Elu[1,*], Matthias Baudisch[1], Hugo Pires[1,†], Francesco Tani[2], Michael H. Frosz[2], Felix Köttig[2], Alexey Ermolov[2], Philip St.J. Russell[2], Jens Biegert[1,3]

[1]ICFO - Institut de Ciencies Fotoniques, The Barcelona Institute of Science and Technology, 08860 Castelldefels (Barcelona), Spain
[2]Max-Planck Institute for Science of Light, Staudtstraße 2, 91058 Erlangen, Germany
[3]ICREA, Pg. Lluís Companys 23, 08010 Barcelona, Spain

*Corresponding author: ugaitz.elu@icfo.eu

†Present address: GoLP/Instituto de Plasmas e Fusão Nuclear, Instituto Superior Técnico, 1049-001 Lisboa, Portugal

**We report a 21-W mid-IR OPCPA that generates 131-µJ and 97 fs (sub-9-cycle) pulses at 160 kHz repetition rate and at a centre wavelength of 3.25 µm. Pulse-to-pulse stability of the CEP-stable output is excellent with 0.33% rms over 288 million pulses (30 min) and compression close to a single optical cycle was achieved through soliton self-compression inside a gas-filled mid-IR anti-resonant-guiding photonic crystal fibre. Without any additional compression device, stable generation of 14.5 fs (1.35-optical-cycle) pulses was achieved at an average power of 9.6 W. The resulting peak power of 3.9 GW in combination with the near-single-cycle duration and intrinsic CEP stability, make our OPCPA a key-enabling technology for the next generation of extreme photonics, strong-field attosecond research and coherent X-ray science.**

## 1. Introduction

The prospects offered by strong field physics and attosecond science (*1*, *2*) have been a key motivation for the development of long wavelength sources of intense ultrashort pulses (*3*, *4*). Utilising the wavelength-scaling of strong field electron recollision for high harmonic generation (*5*, *6*) has made possible the generation of coherent X-ray radiation in the water window (*7*, *8*). Driving HHG with a carrier envelope phase (CEP)-stable sub-2-cycle pulse at 2 micron has recently resulted in the first source of isolated attosecond soft X-ray pulses in this spectral region (*9*), thereby combining attosecond temporal resolution with the element-specificity of X-rays (*10*, *11*). An important demand on laser drivers for strong field physics is hence the capacity to counteract the unfavourable long-wavelength scaling of the underlying processes that results in very low event rates in measurements. Examples are the photoionization with mid-IR radiation (*3*, *4*) or attosecond X-ray generation. Therefore, our near-decade long development of ultrafast mid-IR sources (*12*, *13*) aimed at balancing repetition rate against achievable pulse energy. Our development of a 160-kHz mid-IR OPCPA, which could reach focused intensities of $10^{14}$ W/cm², was instrumental to develop laser-induced electron diffraction (LIED) (*14–17*) to image bond breaking in a single polyatomic molecule with 6 pm spatial and 0.6 fs temporal resolution (*18*).

The combination of ultrafast high peak power and average power represents an important frontier in laser science itself, quite apart from the importance of achieving a measured balance between high repetition rate and high pulse energy in strong field experiments. Ultimately, these requirements present a tremendous challenge for, e.g., OPCPA development, due to material imperfections, nonlinear effects and thermal issues (*19*). As a result, only few sources exist that are capable to delivering ultrashort mid-IR pulses at micro-joule pulse energies (*12*, *20–22*) together with CEP stability (*12*, *23*). In previous works, we have shown that filamentation of pulses in the mid-IR (*24*), i.e. at 3 micron and above, can lead to efficient self-compression by exploiting the interplay of linear and nonlinear propagation effects in the anomalous dispersion regime (*25*) to reach the few-cycle regime. Employing this method, self-compression was demonstrated in solids (*25*, *26*) as well as in gases through self-guiding (*27*) and in hollow-core waveguides (*28*). Common to all these demonstrations is the fact that, despite the vastly varying parameters of the systems used, the average compressed output power always ended up in the milliwatt-level power regime.

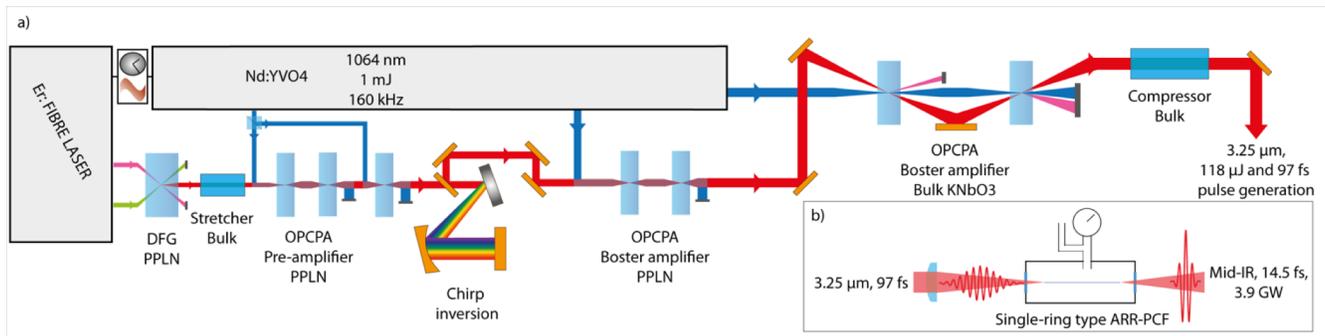

Fig. 1. Setup of high-power, mid-IR OPCPA system. a) The seed is generated by a two-colour fibre front-end in combination with a DFG stage. Afterwards, the mid-IR pulses are stretched and consecutively amplified in a pre-amplifier and two booster-amplifiers. Maximum conversion efficiencies are achieved by multiple use of the pump beam and by individually tailored seed-to-pump pulse durations. The mid-IR output is compressed in a bulk stretcher and b) the final compression to a single optical cycle is performed using an argon-filled ARR-PCF.

Here, we report a mid-IR OPCPA and the compression of its output in a gas-filled hollow core anti-resonant-reflection photonic crystal fibre (ARR-PCF), extending in this way the existing parameters by more than an order of magnitude to achieve peak powers of 3.9 GW at 160 kHz repetition rate with intrinsically CEP-stable pulses near the single-cycle limit.

## 2. Mid-IR OPCPA at 21 W

Figure 1 shows the conceptual layout. The CEP-stable and broadband mid-IR seed was generated, identical to our previous system (*29*), from a two-colour fibre laser (Toptica Photonics AG) by means of difference frequency generation (DFG) (*30*). The Er-fibre system had two parallel outputs of sub-100 fs pulses with 220 mW power at a centre wavelength of 1550 nm and with 100 MHz repetition rate. One of these outputs was frequency down-shifted in a photonic crystal fibre to 1050 nm, thereby yielding two phase coherent pulses which originated from a common pulse, but with different centre wavelength. Difference frequency generation between those two pulses, at 1550 nm and 1050 nm, in a 1-mm-thin periodically poled lithium niobate (PPLN) crystal resulted in a CEP-stable mid-IR pulse with 7 pJ energy; the spectrum is shown in Fig. 2. After propagation through 50 mm of sapphire, the seed pulse was negatively stretched to a duration of 3 ps. As pump, we used a neodymium-doped vanadate (Nd:YVO$_4$) based master oscillator power amplifier (MOPA) from Coherent Inc. providing 1.1-mJ 9-ps pulses at 1064 nm wavelength and 160 kHz repetition rate. The temporal overlap between the pump and the seed was ensured via electronic synchronization between the two lasers to a level better than 300 fs.

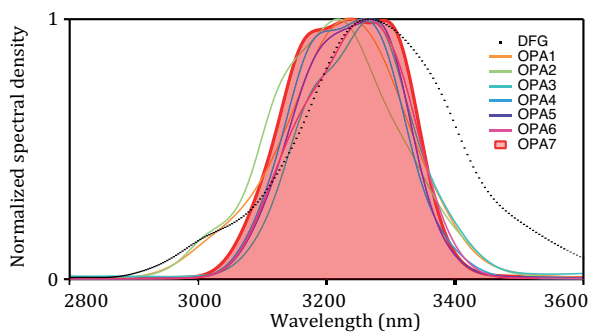

Fig. 2. The spectrum of the mid-IR seed from the DFG (dashed black curve) together with the spectra from each OPA stage (solid curves) and the final output (filled and solid red curve).

The OPCPA comprised of a chain of OPA stages including a pre-amplification section, with three consecutive OPA stages, and a booster section, with four OPA stages. The three OPA stages in the pre-amplification section consisted of one 1.4-mm-long PPLN and two 5-mm-long APPLN crystals (HC Photonics), resulting in pulse energies of 50 nJ, 500 nJ and 2.6 µJ. The first OPA yielded the highest gain ($3.6 \times 10^3$) with the entire pre-amplification stage achieving a cumulative gain of $3 \times 10^5$. The spectral evolution throughout the system is depicted in Fig. 2.

Before further amplification, we reverse the chirp of the mid-IR seed from negative to positive with a grating-based, Martinez-type chirp inverter. The chirp inverter served the purpose of giving control over the chirp and pulse duration within the OPCPA chain after the high gain section and it permitted using bulk sapphire for compression, thus resulting in the highest possible compression efficiency.

For the booster section, we increased the seed pulse duration to 7 ps which maximized the temporal overlap with the pump pulses. Each booster stage consisted of two identical crystals in which the residual pump from the first OPA crystal was recycled for the second OPA crystal, thus increasing the overall efficiency. 2-mm-long PPLN crystals were used for the first two OPAs (OPA4 and OPA5) which were pumped by 250 µJ (40 W) pulses, resulting in amplification of the mid-IR pulses first to 10 µJ and then to 18 µJ, respectively. Here, and in the following two stages, we were able to counteract the spectral energy loss in the blue part of the spectrum in addition to slight spectral broadening in the red part, by using a slightly detuned second stage. Based on our previous investigation of high power effects in mid-IR amplification of different nonlinear media (*19*), we elected to use 5 x 5 mm$^2$ potassium niobate (KNbO$_3$) crystals for the last two OPAs (OPA6 and OPA7). Both stages employed AR-coated 5 x 5 mm$^2$ KNbO$_3$ crystals, which were cut at 40.5° for a non-collinear interaction with an internal angle of 5.2°, using a combined pump energy of 625 µJ (100 W). We achieved amplification to 77 µJ and 131 µJ, respectively, corresponding to an average power of 21 W. Finally, compression was achieved by double-passing through a 10-cm-long AR-coated sapphire rod. We measured a pulse duration of 97 fs (sub-9 optical cycles) with a pulse-to-pulse stability of 0.33% rms over 30 min (see Fig. 3).

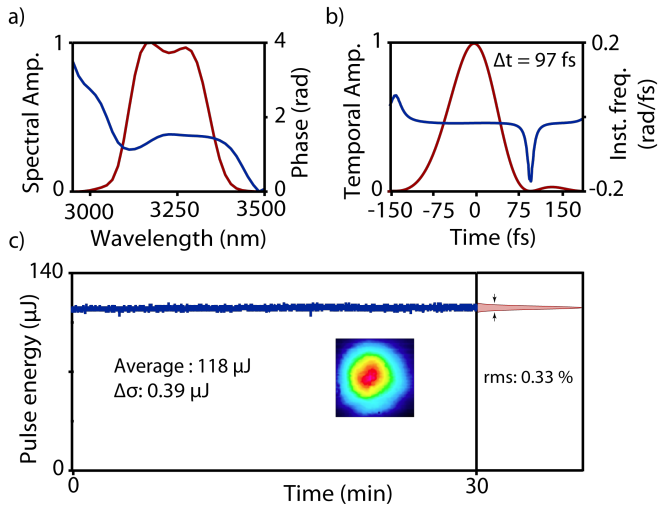

Fig. 3. Output characteristics of the mid-IR OPCPA system. SHG-FROG retrieval of mid-IR output pulses, showing (a) spectral amplitude and phase, and (b) temporal amplitude and instantaneous frequency. (c) Pulse-to-pulse power stability measured over 30 minutes. The inset shows the histogram distribution of the stability measurement and the output beam profile.

## 3. Single-cycle compression in the mid-IR

The conventional scheme of spectral broadening via self-phase modulation (SPM) in a gas-filled capillary followed by dispersion compensation cannot be applied to compress further the output of this system. The capillary loss scales as $\lambda^2/a^3$, where $\lambda$ is the wavelength and $a$ is the radius, while the characteristic length for SPM scales as $L_{SPM} = \lambda/(2\pi n_2 I_0)$, where $n_2$ is the Kerr coefficient and $I_0$ the laser peak intensity in the waveguide. Moreover, a specific challenge for employing gas-filled hollow-core waveguides for the mid-IR is material absorption. This translates into very high propagation loss of mid-IR pulses in Bragg waveguides and fibre capillaries (*31*). To overcome these limitations we employed a new type of ARR-PCF (*32–35*) which consisted of a single ring of anti-resonant capillaries with ~1.2 μm wall thickness, surrounding a 88-μm central hollow core (see Fig. 4).

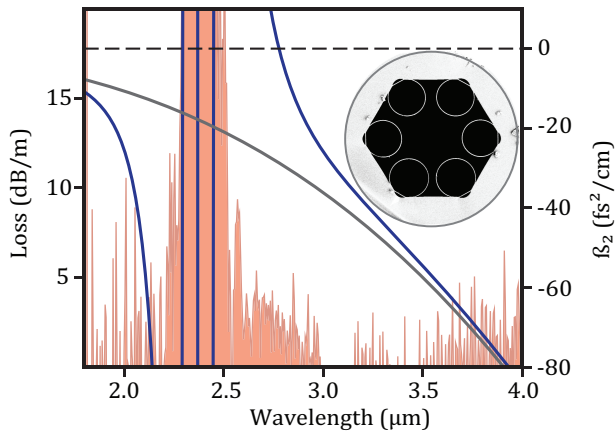

Fig. 4. Measured transmission loss (red) together with the dispersion obtained via the Marcatili model (grey) and fitting FEM calculations (blue) for a 88 μm core evacuated fibre with 1.2 μm core wall thickness. The inset shows a scanning electron microscope (SEM) image of the ARR-PCF fibre.

It has been shown that the ARR-PCF achieves broadband transmission with very low propagation loss in the mid-IR despite being manufactured from fused silica (*34*). Thus, the ARR-PCF presents the enticing possibility for the mid-IR to exploit soliton dynamics for pulse self-compression in the mid-IR by balancing nonlinear propagation with waveguide dispersion (*41*). Soliton self–compression is possible $L_d/L_{SPM}>1$, where $L_d = T_0^2/\beta_2$ is the dispersion length, $T_0$ the input pulse duration and $\beta_2$ is the group velocity dispersion (GVD) (*36–38*). Under such conditions, the cumulated linear dispersion is balanced by nonlinear dispersion, and negative GVD causes new frequency components (generated through self-phase modulation (SPM)) at the leading and trailing parts of the pulse to move towards the pulse centre. Thus, the pulse self-compresses by a factor which is inversely proportional to $N$ upon propagating a distance $\sim L_d/N$ (*37, 39, 40,41*). This distance becomes longer as higher order dispersion terms become more important and for chirped input pulses. As the soliton order increases, the nonlinear chirp introduced by SPM is no longer balanced by negative GVD so that the amount of energy contained in the compressed pulse peak, rather than in its pedestal, decreases. As a result small values of $N$ are required to achieve clean compression (N ≲ 5) (*36, 40*) with good pulse fidelity.

The spectral broadening and concomitant compression depend on the exact dispersion profile and propagation loss of the fibre. The measured transmission loss (Fig. 4) reveals a loss band at 2.4 μm, caused by an anti-crossing between the core mode and the first order resonance in the capillary walls (*42*), which alters the dispersion landscape. Together with the absence of accurately tabulated gas properties from 2.4 μm (*43*) into the mid-IR, this meant that we had to resort to changing the gas species, pressure and fibre length so as to optimize the performance. The ARR-PCF was set up inside a small gas cell with 3-mm-thick $CaF_2$ input and exit windows and the fibre entrance was placed at the image plane of a diamond pinhole. This setup reduced the available pulse energy to 75 μJ (12 W), but it prevented possible damage of the ARR-PCF at high power through coupling mismatch. The light was coupled into the fibre using an achromatic lens with 75 mm focal length and the emerging light was re-collimated with a 100-mm focal length parabolic mirror. We recorded the spectra with a Fourier-transform infrared spectrometer (FTIR) and retrieved the temporal waveforms from SH FROG with the time axis ambiguity removed. The SH employed all-reflective optics and a 30-μm-thin GaSe crystal for sum frequency mixing.

Figure 5 shows a summary of measured pulse durations (measured with SH FROG) for various noble gases, fibre lengths and pressures.

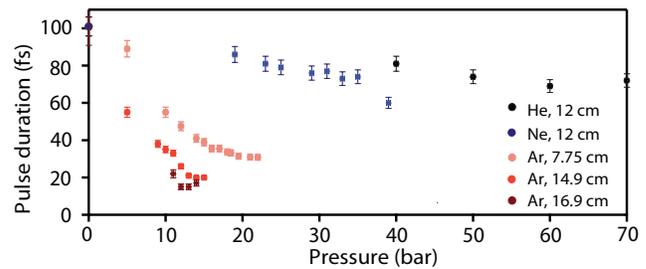

Fig. 5. Measured self–compressed FWHM pulse duration of mid-IR pulses in the ARR-PCF for different noble gas atmospheres and pressures.

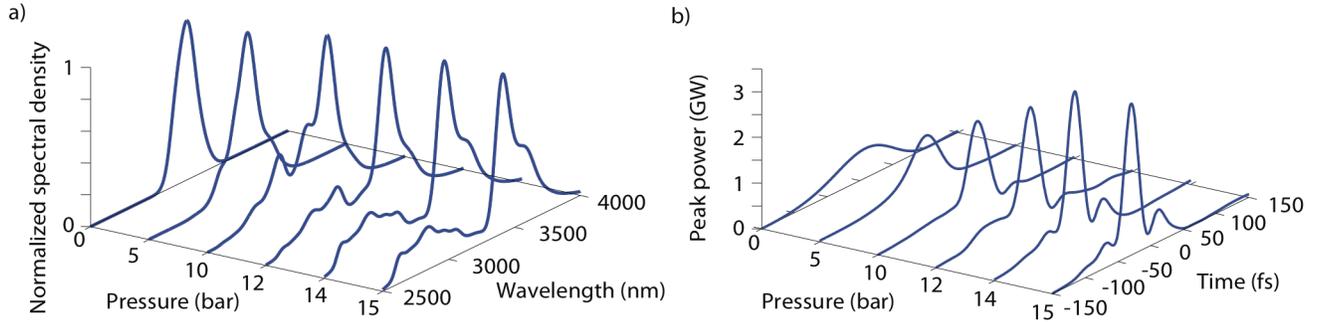

Fig. 6. (a) Spectra measured at the output of a 14.9 cm long length of fibre and (b) the corresponding temporal profile. Both are shown as function of the Ar pressure inside the fibre. The temporal information was obtained by retrieving the traces measured with the SHG FROG.

The parameter scan (Fig. 5) shows that gases with high ionization potential and lower nonlinearity, such as helium (24.6 eV) and neon (21.6 eV), were found not to produce significant self-compression, even at pressures up to 70 bar. We attribute this to the change in dispersion caused by the anti-crossing at 2.4 µm. Decreasing the ionization potential and increasing the gas nonlinearity by choosing argon (15.8 eV), resulted in significant compression to the few-cycle regime. Figure 6 shows the spectral evolution in a 14.9 cm long fibre for vacuum and for up to 15 bar of argon. Clearly visible is the onset of spectral broadening at pressures above 5 bar, and expanding from 2.5 µm to 4 µm for pressures higher than 10 bar. We found, in good agreement with the measurement shown in Fig. 4, that spectral broadening could be extended to 2.4 µm but not further due to resonant absorption within the ARR-PCF.

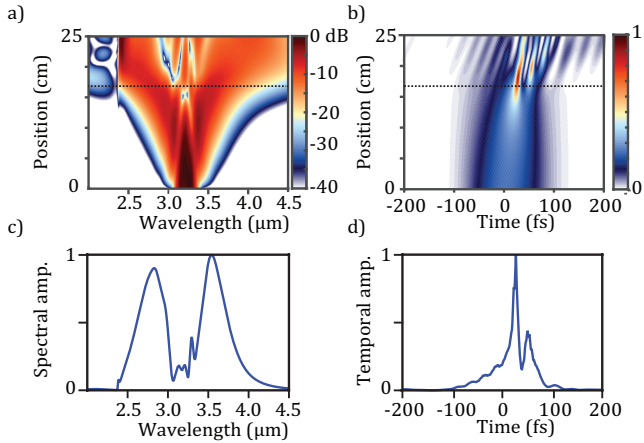

Fig. 7. (a) Simulated spectral and (b) temporal evolution of the experimental input pulse along a 25-cm-long ARR-PCF filled with 12 bar Ar. The dashed lines show the experimental fibre length of 16.9 cm. (c) The spectral and (d) the temporal profiles for the fibre length of 16.9-cm-long ARR-PCF at the fibre output.

Based on this information, we modelled the optimal condition for pulse compression in the ARR-PCF with a unidirectional full field equation code (*44*), including photoionization with the ADK ionisation rate (*45*). As there are no reported measurements for the nonlinear refractive index $n_2$ at 3.3 µm, we used the value measured at 1.05 µm (*46*). To include the contribution of the loss band to the dispersion, we used the effective refractive index obtained through a fit of the results of FEM calculations. We found that in the absence of the loss band at ~2.4 µm, pulses with single cycle duration should be achieved upon propagation in a 12-cm-long fibre filled with ~30 bar He. Because of the steep change in $\beta_2$ associated with the loss band, however, the self-compression dynamics are delayed and a medium with larger nonlinearity is needed to balance the linear dispersion of the fibre. The plot in Fig. 5 supports this as it shows that filling the fibre with Ar rather than He or Ne yields much shorter pulses, which approach the single cycle. Searching for conditions for best compression, the simulation predicted the generation of 14.3-fs pulses after 16.9 cm of propagation at 12 bar Ar (see Fig. 7).

Our experimental findings are in striking agreement with the simulations. The combination of 16.9-cm-long fibre and 12 bar of argon yielded a pulse duration of 14.5 fs which corresponds to 1.35 optical cycles at 3.3 µm centre wavelength. Figure 8 shows the results.

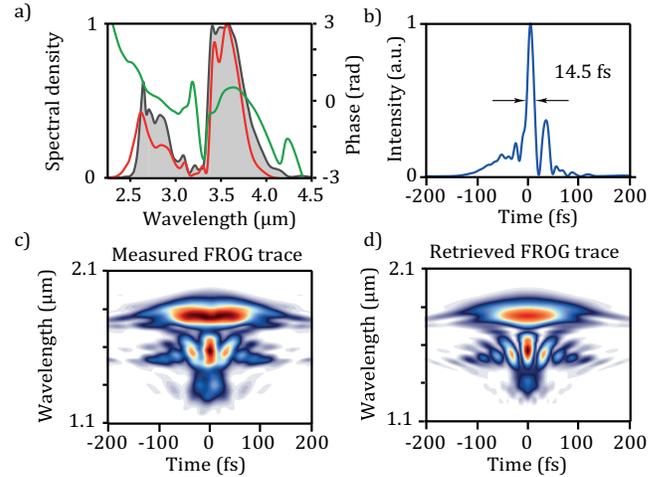

Fig. 8. (a) Measured spectral density (shaded profile), spectral profile (red) and spectral phase (green) retrieved from the SHG FROG trace. (b) Retrieved temporal profile (blue) and retrieved instantaneous frequency (red). (c) Measured and (d) retrieved SHG FROG traces.

We note that the pulse exited the gas cell through a 3-mm-thick $CaF_2$ window and propagated through about 1 m of dry air. We added a 0.5-mm-thick silicon plate to counter-balance most of the accumulated dispersion acquired from propagation through the $CaF_2$ window and through dry air to the GaSe crystal of the SH-FROG. Numerical back-propagation to the fibre exit resulted in an only slightly increased pulse duration of 17 fs. This clearly proves that soliton self-compression was the mechanism responsible for the reduction to a single cycle of the input duration of 97 fs, and not self-phase modulation with post-compression. This compression scheme was remarkably efficient introducing only 20% loss, thereby yielding 60 µJ output energy, which corresponds to a single-cycle mid-IR pulse with 9.6 W average power and 3.9 GW peak power.

## 4. Summary

In conclusion, we have demonstrated a CEP-stable mid-IR OPCPA with chirp inversion which delivers pulses with 131 µJ energy at 160 kHz repetition rate and 21 W average power. We measured a pulse duration of 97 fs (sub-9 optical cycles) and an excellent pulse–to-pulse stability of 0.33% rms over 30 mins, which corresponds to 288 million consecutive pulses. The laser output was compressed down to 1.35 cycles via soliton self-compression in a gas-filled ARR-PCF, yielding 14.5 fs pulses at 3.3 µm with 9.6 W average power. This unique system combines the highest demonstrated average power together with single-cycle duration in the mid-IR and high peak power (3.9 GW) required for experiments in the strong field regime. Furthermore, the fibre compressor offers the possibility to deliver mid-IR single-cycle pulses directly to a gas jet for high harmonic generation and possibly to control the harmonic spectrum via the soliton dynamics (*47*). This system presents a significant step forward for the generation of coherent hard X-rays and the subsequent access to the zeptosecond regime of light-matter interaction.

**Funding.** Spanish Ministry of Economy and Competitiveness "Severo Ochoa" Programme for Centres of Excellence in R&D (SEV-2015-0522), FIS2014-56774-R, Catalan Agencia de Gestió d'Ajuts Universitaris i de Recerca (AGAUR) SGR 2014-2016, Fundació Cellex Barcelona, CERCA Programme / Generalitat de Catalunya, Laserlab-Europe (EU-H2020 654148).

**Acknowledgment**. We thank Daniel Sanchez and Dr. Tobias Steinle for helpful discussions on mid-IR pulse characterization and Prof. Dr. John Travers for consultations on mid-IR fibres and pulse dynamics.